\documentclass[prl,aps,superscriptaddress,twocolumn,showpacs]{revtex4}
\usepackage{bm}
\usepackage{epsfig}
\usepackage{amsmath}
\usepackage{color}

\renewcommand{\v}[1]{{\bf #1}}
\newcommand{\bpm}{\begin{pmatrix}}
\newcommand{\epm}{\end{pmatrix}}
\newcommand{\ba}{\begin{eqnarray}}
\newcommand{\ea}{\end{eqnarray}}
\newcommand{\nn}{\nonumber \\}
\newcommand{\pd}{p^\dag}
\newcommand{\fd}{f^\dag}
\newcommand{\pr}{\prime}

\begin{document}

\title{Emergent $p$-wave Kondo Coupling in Multi-Orbital Bands
\\ with Mirror Symmetry Breaking}

\author{Jun Won Rhim}
\email[Electronic address:$~~$]{phyruth@gmail.com}
\affiliation{School of Physics, Korea Institute for Advanced Study,
Seoul 130-722, Korea}
\author{Jung Hoon Han}
\affiliation{Department of Physics and BK21 Physics Research
Division, Sungkyunkwan University, Suwon 440-746, Korea}
\affiliation{Asia Pacific Center for Theoretical Physics, Pohang,
Gyeongbuk 790-784, Korea}
\date{\today}


\begin{abstract} We examine Kondo effect in the periodic Anderson model for
which the conduction band is of multi-orbital character and subject
to mirror symmetry breaking field imposed externally. Taking
$p$-orbital-based toy model for analysis, we find the Kondo pairing
symmetry of $p$-wave character emerges self-consistently over some
regions of parameter space and filling factor even though only the
on-site Kondo hybridization is assumed in the microscopic
Hamiltonian. The band structure in the Kondo-hybridized phase
becomes nematic, with only two-fold symmetry, due to the $p$-wave
Kondo coupling. The reduced symmetry should be readily observable in
spectroscopic or transport measurements for heavy fermion system in
a multilayer environment such as successfully grown recently.
\end{abstract}
 \maketitle


\textit{Introduction}.- State-of-the-art technique to grow heavy
fermion compounds with atomic thickness precision has been achieved
in recent years~\cite{matsuda-science,matsuda-NP}. Heavy fermion
phase and superconductivity were identified in CeCoIn$_5$ layer of
less than one nanometer thickness grown in this manner, prompting
speculations of exotic new phases in atomically thin
heterostructure~\cite{matsuda-NP}. Apart from the novelty of reduced
dimensionality, the alternate stacking of heavy fermion layer with
normal metal YbCoIn$_5$ layer realized in the growth process is
likely to give rise to new physical phenomena by virtue of the
mirror symmetry breaking (MSB) at the interface. Aspects of
non-centrosymmetric superconductivity have been exhaustively studied
since its original discovery in the heavy-fermion compound
CePt$_3$Si~\cite{CePtSi}. The heavy fermion multilayer structure, on
the other hand, differs from the conventional non-centrosymmetric
material in that the violation of symmetry is macroscopically
imposed (not microscopically at the level of crystal structure) and,
with recently available techniques such as electrolyte gating,
controllable. Due to the thin layer thickness, such MSB is likely to
infect the entire heavy fermion band structure and therefore the
nature of emerging Kondo and/or superconducting pairing as well.
Some consequences of the MSB in the heavy fermion multilayer have
been studied theoretically by Maruyama \textit{et al}. quite
recently~\cite{sigrist}, where the main focus was on the novel
superconducting pairing symmetries and other thermodynamic
properties induced by the phenomenological Rashba term.

\begin{figure}
\includegraphics[width=0.9\columnwidth]{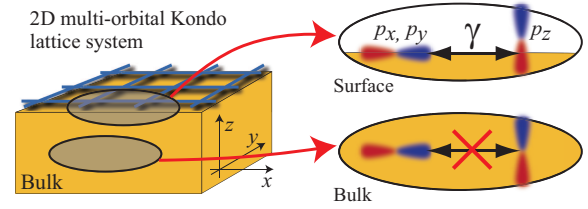}
\caption{(Color online) Schematic picture of Kondo lattice model
subject to mirror symmetry breaking at the surface or interface.
Hopping processes between, for instance, $p_z$ and $p_{x,y}$
otherwise forbidden in the bulk become possible due to the symmetry
breaking field as shown in the upper right
cartoon.}\label{fig:system}
\end{figure}

It has been customary in the past to regard the Rashba interaction
as a relativistic effect manifesting itself in a mirror
symmetry-broken environment. Several recent
papers~\cite{hedegard,OAM1,OAM2}, however, have shown that
multi-orbital degeneracy, in addition to MSB, is an essential
pre-requisite for the manifestation of strong Rashba coupling in the
electron dynamics. In this new picture Rashba interaction becomes an
electrostatic effect with its magnitude dictated by the degree of
orbital asymmetry about the mirror plane (see Fig.
\ref{fig:system})~\cite{hedegard,OAM1,OAM2}.

In fact Kondo lattice behavior takes place in an orbital-rich
environment with the extended $p$- or $d$-orbital states forming the
conduction bands while the localized moments arise from
$f$-orbitals~\cite{hasegawa,cox}. Using our recent experience with
the Rashba phenomena in non-Kondo materials as a guide, we present a
faithful modeling of Kondo lattice behavior where multi-orbital
degrees of freedom in a mirror symmetry-broken environment are
explicitly embedded in the conduction band Hamiltonian. The use of
Rashba interaction as a purely phenomenological addition, as is
often done in the current literature, is avoided. Our approach
differs as well from the large-$N$ generalization of the one-band
Kondo lattice model, where $N$ identical copies with the same
orbital symmetry are adopted as a way to justify the mean-field
approximation~\cite{cox}. In reality the MSB field - usually the
electric field acting perpendicular to the interface or the surface
- influences orbitals which are extended along the field direction
more than those which extend in the orthogonal direction.
Conventional one-band or large-$N$ approach assuming featureless
orbitals will not capture this effect properly.
\\


\textit{The model.}- Choosing the degenerate $p$-orbital states
hopping on a square lattice as a toy example, we write down the
Hamiltonian~\cite{OAM1,OAM2}

\begin{align}
H_{\mathrm{MSB}}=&\sum_{\v k \lambda\sigma}
\left(\epsilon^{\lambda}_{\v k }p^\dag_{\lambda\v k
\sigma}p_{\lambda\v k \sigma}+3\gamma i s_\lambda p^\dag_{z\v k
\sigma}p_{\lambda\v k \sigma}+\mathbf{h.c.}\right)\nn
&~~~~~~ +\sum_{\v k \sigma}\mu_f f^\dag_{\v k \sigma}f_{\v k \sigma}
\label{eq:H-MSB}
\end{align}
for the dispersive $p$-orbital ($p_{\lambda \v k \sigma}$) bands and
a single pair of localized Kramers doublet ($f_{\v k \sigma}$) with
on-site energy $\mu_f$. The orbital index $\lambda\in\{x,y,z\}$
spans the three $p$-orbital states of the conduction band and
$\sigma=+(-)$ is the spin up(down). One can show, employing the two
Slater-Koster parameters $V_1$, $V_2$ for $\sigma$- and
$\pi$-bonding of $p$-orbitals, that $\epsilon^{\lambda}_{\v k
}=-2(V_1+V_2)c_\lambda+2V_2(c_x+c_y)$ with $c_{x(y)} = \cos k_{x(y)}$,
$s_{x(y)} = \sin k_{x(y)}$ and $c_z=s_z=0$ for the assumed
two-dimensional bands. Mirror symmetry breaking is reflected in the
second term of Eq. (\ref{eq:H-MSB}), proportional to the MSB
parmaeter $\gamma$~\cite{OAM2}. They give the hybridization of
$p_z$-orbital with $p_x$- and $p_y$-orbitals; these are processes
normally forbidden by symmetry if a mirror plane existed. This kind
of symmetry breaking is realized naturally on the surface of a bulk
material or by applying vertical electric field to the planar system
as illustrated in Fig. \ref{fig:system}.

Secondly the Kondo hybridization of local and itinerant spins
follows from the Heisenberg-type exchange interaction $H_\mathrm{K}
= J_\mathrm{K}\sum_{i, \lambda} \v S_{i\lambda} \cdot \v S_{if}$,
with spin-1/2 operators made out of $p_\lambda$-orbital ($\v
S_{i\lambda }$) and $f$-orbital ($\v S_{if}$), respectively. Finally
$H_{\mathrm{SOC}}=( \alpha/2 )\sum_i \v L_i \cdot \bm \sigma_i$
involving the $p$-orbital angular momentum operator ($L=1$) $\v L_i$
and the spin-1/2 operator $(1/2)\bm \sigma_i$ at each site $i$ gives
the spin-orbit coupling (SOC) within the $p$-orbital bands. In
typical heavy fermion matter such as CeCoIn$_5$ the heavy element In
playing a main role in the conduction band structure will also carry
substantial amount of spin-orbit energy $\alpha$.

The total Hamiltonian
$H=H_{\mathrm{MSB}}+H_{\mathrm{SOC}}+H_{\mathrm{K}}$ can be handled
within the framework of renormalized mean-field
theory~\cite{senthil-1,senthil-2,vojta} which decomposes the Kondo
exchange interaction as ($\overline{\sigma} = -\sigma$)

\begin{align}
H_{\mathrm{K}}=&\frac{J_\mathrm{K}}{2}\sum_{i\lambda\sigma}
\left(\frac{b^{\lambda}_{\bar{\sigma}\sigma}}{2} \pd_{\lambda
i\sigma} f_{i\bar{\sigma}}-h^{\lambda}_{\sigma} \pd_{\lambda
i\sigma} f_{i\sigma}+\mathbf{h.c.}\right)\nn
&+\frac{J_\mathrm{K}}{2} \sum_{i\sigma}\Gamma^{p}_{\sigma}
\fd_{i\bar{\sigma}} f_{i\sigma}+
\frac{J_\mathrm{K}}{2}\sum_{i\lambda\sigma}
\Gamma^{f}_{\sigma}\pd_{\lambda i\bar{\sigma}} p_{\lambda i\sigma} .
\label{eq:MF_Ham_1}
\end{align}
In addition to the Kondo hybridization parameters
$b^{\lambda}_{\sigma\sigma^\pr}=\langle\fd_{i\sigma} p_{\lambda
i\sigma^\pr}\rangle$ and
$h^{\lambda}_{\sigma}=b^{\lambda}_{\sigma\sigma}+
b^{\lambda}_{\bar{\sigma}\bar{\sigma}}/2$, we also allow
same-orbital spin-flip processes encoded by
$\Gamma^{p}_{\sigma}=\sum_{\lambda} \langle\pd_{\lambda i\sigma}
p_{\lambda i\bar{\sigma}}\rangle$ and
$\Gamma^{f}_{\sigma}=\langle\fd_{i\bar{\sigma}} f_{i\sigma}\rangle$.
The spin-flip parameters $\Gamma^{p}_{\sigma}$,
$\Gamma^{f}_{\sigma}$, and $b^\lambda_{\sigma\bar{\sigma}}$ can be
nonzero due to the spin-orbit interaction. We restrict ourselves
however to non-magnetic, singlet Kondo coupling in the mean-field
scheme and avoid the possibility of Kondo pairing between remote
sites~\cite{senthil-1}, which is another known avenue to obtain
non-$s$-wave Kondo pairing. The fact that our model can exhibit the
$p$-wave Kondo phase despite the on-site-only Kondo hybridization
Hamiltonian $H_\mathrm{K}$ emphasizes the emergent nature of
unconventional pairing. As demonstrated in Refs.
\cite{hedegard,OAM1,OAM2}, Rashba interaction is another emergent
property of the band that can be derived from the underlying
microscopic Hamiltonian consisting only of $H_\mathrm{MSB}$ and
$H_\mathrm{SOC}$.
\\

\begin{figure}
\includegraphics[width=0.9\columnwidth]{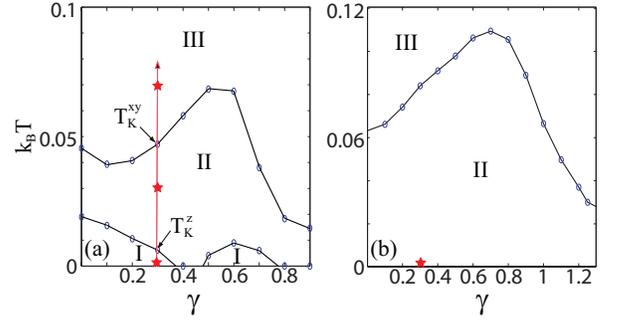}
\caption{(Color online) Phase diagrams with (a) $\alpha =0$ (no
SOC), $n_c =2$ and (b) $\alpha =0.5$ (SOC), $n_c = 1.5$. Phase I
denotes all three $p$-orbitals hybridized ($h^x_\sigma, h^y_\sigma,
h^z_\sigma\neq 0$). Phase II represents only $p_x$, $p_y$ orbitals
hybridized ($h^x_\sigma \neq 0, h^y_\sigma \neq 0$, $h^z_\sigma=0$).
Phase III is trivial with no Kondo hybridization ($h^\lambda_\sigma
= 0$). In (a), as shown by a red vertical arrow, we may encounter
two distinct Kondo temperatures labeled by $T_\mathrm{K}^z$ and
$T_\mathrm{K}^{xy}$. Parameters used to obtain both phase diagrams
are $(V_1, V_2, J_\mathrm{K}) =( 2, 0.65, 3)$. Fermi surface
structures at the four points indicated by stars are shown in Fig.
\ref{fig:bands-2D}. }\label{fig:phase-diagram}
\end{figure}


\textit{Phase diagram.}- All the hybridization parameters can be
determined self-consistently, along with $\mu_f$ to ensure $\langle
\sum_\sigma f^\dag_{i\sigma} f_{i\sigma} \rangle =1$, and the
overall chemical potential $\mu$, for various input band structure
parameters $V_1, V_2, \gamma$, electron filling factor $n_c =
\langle \sum_{\lambda\sigma} p^\dag_{\lambda i \sigma} p_{\lambda i
\sigma}\rangle$, at various temperatures $T$. Generic phase diagrams
resulting from our model Hamiltonian are plotted in Fig.
\ref{fig:phase-diagram} in the space of temperature $T$ and MSB
parameter $\gamma$ with and without the SOC. We observe two distinct
Kondo-coupled phases labeled I in Fig. \ref{fig:phase-diagram},
where all three $p$-orbitals hybridize with $f$-orbital
($h^x_\sigma, h^y_\sigma, h^z_\sigma \neq 0$), and II where only the
in-plane $p_x, p_y$ orbitals hybridize ($h^z_\sigma=0$). The
high-temperature non-Kondo phase with $h^\lambda_\sigma=0$ is
labeled III. Both transitions, I$\rightarrow$II and
II$\rightarrow$III, are second-order. Phase I becomes unstable and
gives way to phase II as shown in Fig. \ref{fig:phase-diagram}(b) as
the strength of SOC is increased. Although the addition of SOC
complicates the microscopic Hamiltonian, fortunately, many of the
hybridization parameters were found to be vanishing
$\Gamma^{p}_{\sigma}=\Gamma^{f}_{\sigma}=b^z_{\sigma\sigma}=
b^x_{\sigma\bar{\sigma}}=b^y_{\sigma\bar{\sigma}}=0$ even with
finite SOC. Features of the phase diagram are dependent on the
filling factor $n_c$ of $p$-electrons. Phase I is observed within a
small range of $n_c$ and, as indicated in Fig.
\ref{fig:phase-diagram}(a), may even exhibit some re-entrant
behavior with increasing $\gamma$. While the details regarding the
extent of each Kondo phase or the Kondo temperature are sensitive to
the occupation $n_c$, the effect of SOC on them was found to be much
less significant by comparison.
\\

\begin{figure}
\includegraphics[width=0.9\columnwidth]{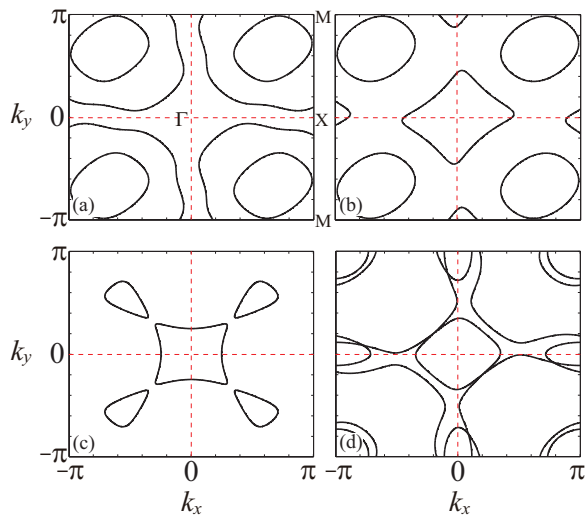}
\caption{(Color online) Band dispersions at the Fermi level for the
various phases marked by red stars ($\star$) in Fig.
\ref{fig:phase-diagram}. (a), (b) and (c) are the Fermi surfaces of
the phases I, II and III in Fig. \ref{fig:phase-diagram}(a)
respectively, and (d) corresponds to the star in Fig.
\ref{fig:phase-diagram}(b). The reduced symmetry in the
Kondo-coupled phases (a), (b), and (d) due to the $p$-wave Kondo
pairing are apparent against the $(k_x, k_y)$ axes in red dashed
lines.}\label{fig:bands-2D}
\end{figure}


\textit{Effective Hamiltonian for p-wave Kondo pairing.}- Most
interestingly, the two Kondo-coupled phases I and II exhibit the
band dispersion with the loss of symmetry under the reflection
$(k_x,k_y) \rightarrow (-k_x,k_y)$ or $(k_x,k_y) \rightarrow
(k_x,-k_y)$, as well as the 90$^\circ$ rotation, as shown in Fig.
\ref{fig:bands-2D}. Reduction in the band structure symmetry only
occurs in the Kondo phase implying that it can only be the
consequence of $p$-wave Kondo coupling. Below we show rigorously how
such unconventional pairing occurs for MSB-infected bands.

With a proper choice of tight binding parameters and filling factors
we can identify cases in the self-consistent band structure wherein
the two lowest-energy spin-orbit coupled $p$-orbital bands and the
two Kramers doublet $f$-electron bands occur in the vicinity of the
Fermi energy while all others are further removed from it.
In such cases one can describe the Kondo physics within the projected
Hilbert space of those four bands associated with the following
operators:

\begin{align}
\tilde{c}_{n\v k} = \sum_{\lambda\sigma}v^{n}_{\lambda\v k
\sigma}p_{\lambda\v k \sigma} ~~\mathrm{and}~~\tilde{f}_{\v k
\sigma} = f_{\v k \sigma} + \sum_{n=3}^6 u^n_{\v k
\sigma}\tilde{c}_{n\v k} .
\end{align}
Here, $n=1-6$ labels the six $p$-orbital bands of
$H_{\mathrm{MSB}}+H_{\mathrm{SOC}}$, the state at $\v k$ being
denoted $|\v v^n_{\v k}\rangle =\tilde{c}^\dag_{n\v k}|0\rangle$
with the coefficients $v^n_{\lambda\v k \sigma}$. The localized
$f$-electron operator is modified through Kondo hybridization with
the higher energy bands, $u^n_{\v k \sigma}=\langle\v v^n_{\v k}|
H_{\mathrm{K}} |f_{\v k\sigma}\rangle/(\mu_f-\varepsilon^c_{n\v
k})$, $n=3-6$, where $\varepsilon^c_{n\v k}$ is the energy of $n$-th
band of $H_{\mathrm{MSB}}+H_{\mathrm{SOC}}$ and $|f_{\v
k\sigma}\rangle=f^\dag_{\v k \sigma}|0\rangle$. The effective
Hamiltonian within the reduced Hilbert space of $\{|\v v^1_{\v
k}\rangle,|\v v^2_{\v k}\rangle,|\tilde{f}_{\v
k+}\rangle,|\tilde{f}_{\v k-}\rangle \}$, where
$|\tilde{f}_{\v k\sigma}\rangle=\tilde{f}^\dag_{\v k
\sigma}|0\rangle$, is

\begin{align}\label{eq:effective_Ham}
H_{\mathrm{eff}}=&\sum_{n=1,2} \varepsilon^c_{n\v
k}\tilde{c}^\dag_{n\v k} \tilde{c}_{n\v k}+\sum_{\sigma}
\varepsilon^f_{\sigma\v k} \tilde{f}^\dag_{\v k \sigma}\tilde{f}_{\v
k \sigma}\nn
&+\sum_{n=1,2, \sigma}\left( V^{n\sigma}_{\v k}\tilde{c}^\dag_{n\v
k}\tilde{f}_{\v k \sigma} + \mathbf{h.c.}\right) .
\end{align}
The $f$-level dispersion $\varepsilon^f_{\sigma\v k}$ and the
effective Kondo coupling $V^{n\sigma}_{\v k} =\langle\v v^n_{\v k}|H_\mathrm{K} |f_{\v
k\sigma}\rangle$ are given by

\begin{align}
\varepsilon^f_{\sigma\v k} &=\mu_f+\sum_{n=3}^6 \frac{|\langle\v
v^n_{\v k}| H_{\mathrm{K}} |f_{\v
k\sigma}\rangle|^2}{\mu_f\!-\!\varepsilon^c_{n\v k}}, \nn
V^{n\sigma}_{\v k} &=\frac{J_\mathrm{K}}{2}
\left(\frac{1}{2}b^z_{\sigma\bar{\sigma}}v^n_{z\v k \bar{\sigma}}
-h^x_{\bar{\sigma}}v^n_{x\v k\sigma}-h^y_{\bar{\sigma}}v^n_{y\v k\sigma}\right).
\label{eq:effective_Kondo}
\end{align}
Some properties of the self-consistently obtained hybridization
parameters of Kondo-coupled phases such as
$\Gamma^{p}_{\sigma}=\Gamma^{f}_{\sigma}=
b^z_{\sigma\sigma}=b^x_{\sigma\bar{\sigma}}=b^y_{\sigma\bar{\sigma}}=0$
are used to derive the above effective Hamiltonian. The energy
dispersion obtained from $H_\mathrm{eff}$ matches the full
dispersion almost exactly. Characteristics of the Kondo-coupled
bands can be discussed faithfully now in terms of the effective
Hamiltonian, $H_\mathrm{eff}$.

\begin{figure}
\includegraphics[width=1\columnwidth]{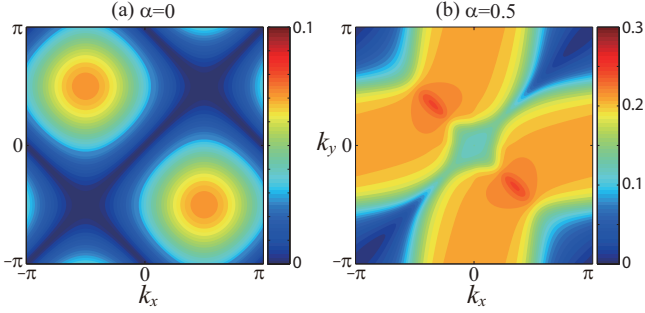}
\caption{(Color online) Contour plots of $|V^{1+}_{\v k}|$, the
absolute value of the effective Kondo hybridization, in the
Brillouin zone. (a) $\alpha=0$ (no SOC) and (b)
$\alpha=0.5$.}\label{fig:kondo-coupling}
\end{figure}

The effective Kondo coupling $V^{n\sigma}_{\v k}$ appearing in
$H_\mathrm{eff}$ has the explicit momentum dependence despite the
fact that the original Hamiltonian only contained the on-site Kondo
interaction. The symmetry of the Kondo pairing, self-consistently
determined to be $p$-wave as illustrated in Fig.
\ref{fig:kondo-coupling}, also dictates that of the hybridized
bands, as shown in Fig. \ref{fig:bands-2D}. The symmetry propertes
of $V^{n\sigma}_{\v k}$, in turn, follow from those of the wave
functions $v^n_{\lambda \v k \sigma}$ as outlined in Table
\ref{table}.

Transformation rules for the wave functions are deduced by examining
whether the Hamiltonians at a pair of symmetry-related momenta can
transform into each other by some kind of unitary or anti-unitary
operators such as $K\sigma_z\otimes R_x(\pi)$, $K I\otimes
R_y(\pi)$, $\sigma_z\otimes R_z(\pi)$ and
$e^{-i\frac{3}{4}\pi}D^{1/2}_z(3\pi/2)\otimes R_z(\pi/2)$. We denote
$R_\lambda(\theta)$ for the SO(3) rotation matrix, $\lambda=x,y,z$,
$D^{1/2}_z(\theta)$ for the SU(2) rotation matrix, and $K$ for the
complex conjugation. From the Table and Eq.
(\ref{eq:effective_Kondo}) one can show that
$V^{n\sigma}_{-k_x,-k_y}=-\sigma V^{n\sigma}_{k_x,k_y}$, and
therefore a unitary transformation $\sigma_z\otimes \mathrm{I}$
between $H_{\mathrm{eff}}(\v k )$ and $H_{\mathrm{eff}}(-\v k )$
exists, ensuring the equivalence of the energies between $\v k$ and
$-\v k$. On the other hand $V^{n\sigma}_{-k_x,k_y}$,
$V^{n\sigma}_{k_x,-k_y}$ and $V^{n\sigma}_{-k_y,k_x}$ are not
related to $V^{n\sigma}_{k_x,k_y}$ in any way according to the Table
\ref{table}. For example, $V^{n\sigma}_{-k_x,k_y} =J_\mathrm{K}
(-b^z_{\sigma\bar{\sigma}}v^{n*}_{z\v k
\bar{\sigma}}/2+h^x_{\bar{\sigma}}v^{n*}_{x\v
k\sigma}-h^y_{\bar{\sigma}}v^{n*}_{y\v k\sigma})/2$ does not even
have the same magnitude as $V^{n,\sigma}_{\v k}$ in general because
the wave function $v^n_{\lambda \v k \sigma}$ are complex-valued.
Referring to Eq. (\ref{eq:effective_Kondo}), neither the reflection
about the $x$- or the $y$-axis nor the 90$^\circ$ rotation in $\v
k$-space is guaranteed to yield unitary-equivalent $H_\mathrm{eff}$.

\begin{table}
\begin{tabular}{|c|c|}\hline
$\v k$ & \rm{wave function} \\
\hline $(k_x,k_y)$ & $\{v^n_{x\v
k +},v^n_{y\v k +},v^n_{z\v k +},v^n_{x\v k -},v^n_{y\v k
-},v^n_{z\v k -}\}$\\\hline
$(-k_x,k_y)$ & $\{-v^{n*}_{x\v k +},v^{n*}_{y\v k
+},-v^{n*}_{z\v k +},-v^{n*}_{x\v k
-},v^{n*}_{y\v k -},-v^{n*}_{z\v k
-}\}$\\\hline
$(k_x,-k_y)$ & $\{v^{n*}_{x\v k +},-v^{n*}_{y\v k
+},-v^{n*}_{z\v k +},-v^{n*}_{x\v k
-},v^{n*}_{y\v k -},v^{n*}_{z\v k
-}\}$\\\hline
$(-k_x,-k_y)$ & $\{-v^n_{x\v k +},-v^n_{y\v k
+},v^n_{z\v k +},v^n_{x\v k -},v^n_{y\v k
-},-v^n_{z\v k -}\}$\\\hline
$(-k_y,k_x)$ & $\{-v^n_{y\v k +},v^n_{x\v k
+},v^n_{z\v k +},-iv^n_{y\v k -},iv^n_{x\v k
-},iv^n_{z\v k -}\}$\\\hline
\end{tabular}
\caption{Relation between eigenstates of
$H_{\mathrm{MSB}}+H_\mathrm{SOC}$ at several symmetry-related
momenta. $v^n_{\lambda\v k \sigma}$ in the first row refers to the
amplitude of $p_\lambda$-orbital with spin $\sigma$ for the
eigenstate at $\v k=(k_x,k_y)$; $\sigma=+(-)$ corresponds to spin
up(down). Eigenstates at other momenta are related to this one as
shown in the subsequent rows.} \label{table}
\end{table}



It is further possible to find an explicit expression of the
effective Kondo coupling to prove its $p$-wave nature for models
without SOC. Given the parameters such as
$V_1=2,~V_2=0.65,~\gamma=0.3$ and the filling of conduction
electrons $n_c=1.5$, the band dispersions around the M point meet
the conditions assumed in deriving the effective Hamiltonian
$H_{\mathrm{eff}}$. Here one can show $v^n_{\lambda \v k \sigma}$
near $\v k = (\pi,\pi)$ is $v^1_{x,y \v k +}\approx-3\gamma
s_{x,y}/2(V_1+V_2)$ and $v^1_{x,y \v k -}=0$, and vice versa for
$n=2$. Turning off SOC, all spin-mixing mean field parameters
$b^z_{\sigma\bar{\sigma}}$ vanish. We can conclude only two
equivalent ground states with real-valued Kondo parameters
$h^x_\sigma =  h^y_\sigma$ or $h^x_\sigma =-h^y_\sigma$ are
possible, giving out

\begin{align}
V_{\v k}\equiv V^{1+}_{\v k}=V^{2-}_{\v
k}=\frac{3iJ_{\mathrm{K}}h^x\gamma}{4(V_1+V_2)}(\sin k_x \pm \sin k_y
) \label{eq:effective_Kondo_2}\end{align}
and $V^{1-}_{\v k}=V^{2+}_{\v k}=0$ for $h^x_\sigma =\pm
h^y_\sigma$, respectively. The effective Kondo coupling indeed has
the $p$-wave symmetry with a node along the $k_x= +k_y$ or
$k_x=-k_y$ line. Figure \ref{fig:kondo-coupling}(b) shows the plot
of $|V^{1+}_{\v k}|$ for the case of $h^x_\sigma =-h^y_\sigma$ where
the node extends along $k_x=-k_y$. It is also seen that $p$-wave
Kondo coupling is not predicated on SOC at all.

Instead, the $p$-wave Kondo coupling is strictly a consequence of
nonzero MSB, $\gamma \neq 0$. When $\gamma=0$, Hamiltonian matrices
of $H_{\mathrm{MSB}}+H_{\mathrm{SOC}}$ at $(k_x,k_y)$, $(-k_x,k_y)$,
$(k_x,-k_y)$ and $(-k_x,-k_y)$ are equal to each
other~\cite{comment}. As a result, those momenta have identical
eigenvectors and, according to Eq. (\ref{eq:effective_Kondo}), the
same Kondo coupling $V^{n\sigma}_{\v k}$.  In addition the $p$-wave
character arises only when both $h^x_{\sigma}$ and $h^y_{\sigma}$
are nonzero. For phases I and II such condition is indeed achieved
with $|h^x_\sigma|= |h^y_\sigma|$. The $p$-wave Kondo states are
favored away from half-filling regime such as considered in the
phase diagram of Fig. \ref{fig:phase-diagram}. At half filling $n_c
=3$ without SOC, on the other hand, we often find the ground phase
with only $h^z_\sigma$ non-vanishing. In this case, the effective
Kondo coupling is in the form $V^{n\sigma}_{\v
k}=-J_{\mathrm{K}}h^z_{\bar{\sigma}}v^n_{z\v k\sigma}/2 \approx
iJ_{\mathrm{K}}h^z_{\bar{\sigma}}/2$, which is $s$-wave as in the
ordinary heavy fermion matter.
\\

\textit{Discussion.}- As the angstrom-scale deposition technique of
thin films develops into maturity, the role of surface states with
broken mirror symmetry takes on greater practical significance. Such
MSB field effect leads to interesting consequences when acting on a
multi-orbital environment such as the emergence of Rashba
effect~\cite{OAM1,OAM2}, orbital-dependent spin transfer
torque~\cite{STT}, and as argued in this work, of un-conventional
Kondo pairing. Conventional one-band Kondo model with the
phenomenological Rashba term could very well miss this feature due
to the inadequate treatment of the multi-orbital character. The
asymmetric band structure we found in the Kondo-coupled phase should
be readily detected in the spectroscopic study such as ARPES, or in
the transport measurements through their directional responses.
Although proposals of non-$s$-wave Kondo pairing are already
available~\cite{senthil-1,senthil-2,vojta}, they tend to trace its
origin to the non-local character of Kondo coupling between
localized and itinerant moments~\cite{senthil-1}, or the orbital
symmetry of the $f$-orbital localized moments~\cite{senthil-2}. By
contrast, the $p$-wave Kondo hybridization emerging in our scenario
is a generic consequence of the macroscopic breaking of mirror
symmetry imposed by the geometry of the heterostructure itself.
Existing theory of heavy fermion heterostructure~\cite{sigrist}
focused on the nature of unconventional superconductivity resulting
from Rashba effect, but did not address the nature of Kondo pairing
itself.

\acknowledgments J. H. H. is supported by NRF grant (No.
2010-0008529, 2011-0015631) and acknowledges discussion with Yuji
Matsuda that inspired this project.


\begin{thebibliography}{24}
\bibitem{comment} One can show that this statement is consistent with Table \ref{table}
after allowing for the double degeneracy of the state due to
vanishing $\gamma$ and the U(1) gauge freedom of eigenvectors.
\end{thebibliography}
\end{document}